\documentclass
[twocolumn,prl,unsortedaddress,superscriptaddress,showpacs]{revtex4}%
\usepackage{amsmath}
\usepackage{amsfonts}
\usepackage{amssymb}
\usepackage{graphicx}
\usepackage[utf8]{inputenc}%
\providecommand{\U}[1]{\protect\rule{.1in}{.1in}}
\providecommand{\U}[1]{\protect\rule{.1in}{.1in}}

\begin{document}
\title{Black Hole Remnants}
\author{Ali H. Chamseddine}
\affiliation{Physics Department, American University of Beirut, Lebanon }
\affiliation{}
\affiliation{Theoretical Physics, Ludwig Maxmillians University, Theresienstr. 37, 80333
Munich, Germany}
\author{Viatcheslav Mukhanov}
\affiliation{Theoretical Physics, Ludwig Maxmillians University, Theresienstr. 37, 80333
Munich, Germany}
\affiliation{MPI for Physics, Foehringer Ring, 6, 80850, Munich, Germany}
\affiliation{School of Physics, Korea Institute for Advanced Study, Seoul 02455, Korea}
\author{Tobias B. Russ}
\affiliation{Theoretical Physics, Ludwig Maxmillians University, Theresienstr. 37, 80333
Munich, Germany}
\date{\today}

\begin{abstract}
We show that in asymptotically free mimetic gravity with limiting curvature
the black hole singularity can be resolved and replaced by a static patch of
de Sitter space. As a result of Hawking evaporation of these non-singular
black holes, there remain stable remnants with vanishing Hawking temperature.

\end{abstract}
\pacs{0.4.20.-q, 0.4.20.Dw, 0.4.70.-s, 0.4.70.Dy}
\maketitle
\preprint{APS/123-QED}

\emph{Introduction.---}
The two unresolved issues of black hole physics actively discussed during the
last 50 years are: \textit{a)} the singularity problem, \textit{b)} the final
state of black hole evaporation and, closely related to it, the information
paradox (see, for instance, \cite{UnruhWald}). Both issues are usually
referred to some, as yet unknown, non-perturbative theory of quantum gravity.
The nature of Planck mass black holes, their stability and the resolution of
the space-like singularity remain unclear. In this letter we suggest a
completely different approach to these problems. Namely, we assume that
classical General Relativity is modified at curvatures close to, \textit{but
still below Planckian curvature}, in such a way that some limiting curvature,
which is a free parameter of the theory, can never be exceeded (cf.
\cite{Markov}). Moreover, we propose that at this limiting curvature the
gravitational constant vanishes. These two assumptions allow us to avoid all
problems related to non-perturbative quantum gravity effects and study in a
fully controllable way the final stage of non-singular evaporating black
holes. To implement the above ideas, we use the mimetic field introduced in
\cite{MC} and further exploited in references \cite{CosmologywithMDM},
\cite{Resolving}, \cite{NonsingularBH}, \cite{massivmim1}, \cite{massivmim2},
\cite{AFmimetic}. We show that in asymptotically free theories with limiting
curvature, black holes generically have stable remnants with mass determined
by the inverse limiting curvature value, thus exceeding Planck mass. These
remnants have vanishing Hawking temperature and, by the arguments shown in
\cite{AFmimetic}, metric quantum fluctuations never become relevant for them.

\emph{The Lema\^{\i}tre coordinates.---}
The metric of both black hole and de Sitter universe in \textquotedblleft
static\textquotedblright\ coordinates can be written as
\begin{equation}
\mathrm{d}s^{2}=\left(  1-a^{2}\left(  r\right)  \right)  \mathrm{d}%
t^{2}-\frac{\mathrm{d}r^{2}}{\left(  1-a^{2}\left(  r\right)  \right)  }%
-r^{2}\mathrm{d}\Omega^{2},\label{1}%
\end{equation}
where $\mathrm{d}\Omega^{2}=\mathrm{d}\vartheta^{2}+\sin^{2}\vartheta
\mathrm{d}\varphi^{2}.$ For a black hole of mass $M$ the function
$a^{2}\left(  r\right)  =r_{g}/r$, where $r_{g}=2M$ and for the de Sitter
universe $a^{2}=\left(  Hr\right)  ^{2}$ with $H^{-1}$ being the radius of
curvature. Throughout this paper we use Planck units where all fundamental
constants are set to unity. In both cases the static coordinate system is
incomplete. Moreover, at the horizon corresponding to $a^{2}=1$, there is a
coordinate singularity. Therefore, in both cases it is more convenient to use
the synchronous Lema\^{\i}tre coordinates%
\begin{equation}
T=t+\int\frac{a}{1-a^{2}}\mathrm{d}r,\text{ \ \ }R=t+\int\frac{\mathrm{d}%
r}{a\left(  1-a^{2}\right)  },\label{2}%
\end{equation}
which are non-singular on the horizons (cf. \cite{Lemaitre}).
In these coordinates metric (\ref{1}) becomes%
\begin{equation}
\mathrm{d}s^{2}=\mathrm{d}T^{2}-a^{2}\left(  x\right)  \mathrm{d}R^{2}%
-b^{2}(x)\mathrm{d}\Omega^{2} ,\label{3}%
\end{equation}
where $a^{2}$ and $b^{2}=r^{2}$ must be expressed in terms of Lema\^{\i}tre
coordinates $T$ and $R$ using the relation
\begin{equation}
x\equiv R-T=\int\frac{\mathrm{d}r}{a\left(  r\right)  },\label{4}%
\end{equation}
which follows from $\left(  \ref{2}\right) $. Note that the norm of the
Killing vector field $\partial/\partial t=\partial/\partial R+\partial
/\partial T$ vanishes wherever $a=1 $. The black hole metric in these new
coordinates becomes%
\begin{equation}
\mathrm{d}s^{2}=\mathrm{d}T^{2}-\left(  x/x_{+}\right)  ^{-2/3}\mathrm{d}%
R^{2}-\left(  x/x_{+}\right)  ^{4/3}r_{g}^{2}\,\mathrm{d}\Omega^{2},\label{5}%
\end{equation}
and it is regular at the horizon $x=x_{+}=4M /3.$ The region $x>0$ covers both
interior and exterior of the black hole and $x=0$ corresponds to the physical
space-like singularity where curvature invariants blow up. Hence, in General
Relativity this metric is not extendable to negative $x.$

Correspondingly, the de Sitter metric takes the form%
\begin{equation}
\mathrm{d}s^{2}=\mathrm{d}T^{2}-\exp\left(  2H(x-x_{-}\right)  )\left(
\mathrm{d}R^{2}+H^{-2}\mathrm{d}\Omega^{2}\right)  ,\label{6}%
\end{equation}
where $x_{-}$ is a constant of integration in $\left(  \ref{4}\right) $ and
the de Sitter horizon occurs at $x=x_{-}.$ The region $x<x_{-}$ corresponds to
the patch of size $r=H^{-1}$ covered by static coordinates, which on larger
scales do not exist.

Calculating the spatial curvature of metric (\ref{3}), for constant $T$
hypersurfaces,  one finds that it vanishes if $a^{2}=\left(  \mathrm{d}%
b/\mathrm{d}x\right)  ^{2}$. Hence, both solutions $\left(  \ref{5}\right)  $
and $\left(  \ref{6}\right)  $ are spatially flat in the Lema\^{\i}tre slicing.

The Schwarzschild metric has a Kasner type space-like singularity. In a
previous paper \cite{AFmimetic} we have found that in a theory where the ideas
of limiting curvature and asymptotic freedom of gravity are realized via the
mimetic field, Kasner singularities are avoided and replaced by a de Sitter
region at limiting curvature. In this paper we implement these same ideas for
black holes. Using ansatz (\ref{3}) for the metric, we will find an explicit
solution describing a non-singular black hole whose metric approaches
asymptotics $\left(  \ref{5}\right)  $ and $\left(  \ref{6}\right)  $ at low
and high curvatures, respectively.

\emph{Modified Mimetic Gravity.---}
Introducing the mimetic field $\phi$ through a Lagrange multiplier constraint,
consider the theory of gravity defined by the action
\begin{equation}
S=\frac{1}{16\pi}\!\int\!\mathrm{d}^{4}x\sqrt{-g}\left(  -\mathcal{L}%
+\lambda\left(  g^{\mu\nu}\phi_{,\mu}\phi_{,\nu}-1\right)  \right)
,\label{action}%
\end{equation}
with Lagrangian density%
\begin{equation}
\mathcal{L}=f(\Box\phi)R+(f(\Box\phi)-1)\,\widetilde{R}+2\Lambda(\Box
\phi),\label{6a}%
\end{equation}
where
\begin{equation}
\widetilde{R}\equiv2\phi^{,\mu}\phi^{,\nu}G_{\mu\nu}-(\Box\phi)^{2}+\phi
^{;\mu\nu}\phi_{;\mu\nu},\label{7}%
\end{equation}
and $G_{\mu\nu}$ is the Einstein tensor. In \cite{AFmimetic} we argued that
$\Box\phi$ is the unique measure of curvature that $f$ can depend on without
introducing higher time derivatives to the modified Einstein equation. The
extension of the action presented in \cite{AFmimetic} by the $\widetilde{R}%
$-term is done with the purpose to remove higher spatial and mixed
derivatives.

Since the Lema\^{\i}tre coordinates are synchronous, the generic solution of
the constraint equation $\phi^{,\alpha}\phi_{,\alpha}=1$ satisfied by the
mimetic field, which is compatible with the symmetries of ansatz (\ref{3}), is
$\phi=T+const.$
In this coordinate system%
\begin{equation}
\Box\phi=\kappa=\frac{\partial}{\partial T}\ln\sqrt{\gamma}=-\frac{\mathrm{d}%
}{\mathrm{d}x}\ln\left(  ab^{2}\right)  \label{7a}%
\end{equation}
represents the trace of extrinsic curvature $\kappa=\kappa_{a}^{a}$ of
synchronous slices at constant $T$. The expression $\widetilde{R}$, given in
(\ref{7}), is nothing but the spatial curvature scalar ${^{3}\!R}$ of these
slices, expressed in covariant form.

Variation of (\ref{action}) with respect to the metric $g_{\mu\nu}$ yields the
modified vacuum Einstein equations. Solving the equation obtained by varying
(\ref{action}) with respect to $\phi$ for the Lagrange multiplier $\lambda$
and substituting the metric ansatz (\ref{3}), after lengthy but
straightforward calculations we find that the spatial components of the
modified Einstein equations have the first integral (see \cite{AFmimetic},
\cite{Note1})%
\begin{equation}
\frac{\dot{b}}{b}-\frac{\dot{a}}{a}=\frac{3M}{fab^{2}},\label{8}%
\end{equation}
where dot denotes the derivative with respect to $x$ and a constant of
integration has been fixed to match the Schwarzschild solution with mass $M$
in the limit $x\rightarrow\infty$. In deriving (\ref{8}) we have assumed that
the spatial curvature remains negligible everywhere for solutions matching the
two spatially flat asymptotics (\ref{5}) and (\ref{6}). Later on we will
justify this assumption.

Accordingly, the temporal modified Einstein equation becomes
\begin{equation}
\frac{\kappa^{2}\left(  f-2\kappa f^{\prime}\right)  -3\left(  \Lambda
-\kappa\Lambda^{\prime}\right)  }{\left(  f+\kappa f^{\prime}\right)
}=\left(  \frac{3M}{fab^{2}}\right)  ^{2},\label{9}%
\end{equation}
where prime denotes the derivative with respect to $\kappa$. Equations
$\left(  \ref{8}\right)  $ and $\left(  \ref{9}\right)  $ determine the two
unknown functions $a\left(  x\right)  $ and $b\left(  x\right)  .$

\emph{Asymptotic Freedom at Limiting Curvature.---}
The inverse running gravitational constant is represented by
\begin{equation}
f\left(  \Box\phi\right)  =\frac{1}{G\left(  \Box\phi\right)  },\label{10}%
\end{equation}
normalized as $f\left(  \Box\phi=0\right)  =1$ in Planck units. Asymptotic
freedom is characterized by a divergence of $f$ as $\Box\phi\rightarrow
\left\vert \kappa_{0}\right\vert $ approaches the limiting curvature
$\kappa_{0}$, which is a free parameter of the theory.
In \cite{AFmimetic} we have shown that in a contracting Kasner universe the
vanishing gravitational constant very efficiently suppresses anisotropies and
the solution close to the limiting curvature becomes isotropic, approaching a
de Sitter universe with $H=\kappa_{0}/3$. Since black hole and Kasner
singularities are similar, one can expect that the black hole singularity can
be resolved in the same way. Namely, the asymptotic solution far away from the
black hole, which is still given by $\left(  \ref{5}\right) $, will start to
approach $\left(  \ref{6}\right)  $ as soon as the curvature approaches its
limiting value. The full solution extends for the entire range $-\infty
<x<+\infty$.

For the Schwarzschild solution ( \ref{5}) the function $a\propto b^{-1/2}$
increases as $b\rightarrow0$, while for the de Sitter solution $a\propto b$.
It follows that $\dot{a}$ has to vanish at some intermediate point $x_{\ast}$
where $a$ reaches its maximum value before starting to decrease as we go
deeper into the black hole. If $a\left(  x_{\ast}\right)  >1$, there exist two
Killing horizons where $a\left(  x_{\pm}\right)  =1$ at $x_{+}>x_{\ast}$ and
$x_{-}<x_{\ast}$, named in analogy with (\ref{5}) and (\ref{6}). In the
limiting case where $a\left(  x_{\ast}\right)  =1$, the two horizons merge at
$x_{+}=x_{-}=x_{\ast}$ and the region where $\partial/\partial t=\partial
/\partial R+\partial/\partial T$ is space-like disappears. This is the case of
a minimal black hole with mass $M_{\min}\sim\kappa_{0}^{-1}$ which exceeds the
Planck mass if the limiting curvature is below the Planckian value. For
$M<M_{\min}$, $a$ is always smaller than unity, no horizon occurs and thus no
black holes with mass smaller than $M_{\min}$ exist.

One can easily find that for the metric (\ref{3}) the surface gravity of the
Killing horizons is given by
\begin{equation}
g_{s}=-\dot{a}(x_{\pm}),\label{11}%
\end{equation}
and it hence vanishes for the minimal black hole. Because the Hawking
radiation temperature is proportional to $g_{s},$ it is equal to zero for
these minimal mass black holes and as a result of evaporation there must
remain stable remnants of mass $M_{\min}.$ Thus, the existence of limiting
curvature combined with asymptotic freedom of gravity generically leads to the
existence of minimal stable black hole remnants.

To demonstrate this in a concrete theory, we will find below, an explicit
spatially flat, exact solution describing a non-singular black hole with
stable remnant. \cite{Note1}

\emph{Exact Solution.---}
Let us take
\begin{equation}
f(\tilde{\kappa} )=\frac{1+3\tilde{\kappa}^{2}}{\left(  1+\tilde{\kappa}%
^{2}\right)  \left(  1-\tilde{\kappa}^{2}\right)  ^{2}},\label{12}%
\end{equation}
where $\tilde{\kappa}\equiv\kappa/\kappa_{0}$ and chose $\Lambda(\kappa) $ in
such a way
that the square root of the branch $\kappa<0$ of (\ref{9}) becomes
\begin{equation}
\frac{-\tilde{\kappa}}{1-\tilde{\kappa}^{4}}=\frac{3M/\kappa_{0}}{ab^{2}%
}.\label{13}%
\end{equation}
Taking the $x$ derivative of the logarithm of this equation and using
(\ref{7a}), we obtain a first order differential equation for $\tilde{\kappa
}\left(  x\right)  $ with the implicit solution
\begin{equation}
-\kappa_{0}\,x=\frac{1}{\tilde{\kappa}}+2\left(  \arctan\tilde{\kappa}%
-\tanh^{-1}\tilde{\kappa}\right)  .\label{14}%
\end{equation}
This provides a one-to-one relation between $x\in(-\infty,\infty)$ and
$\tilde{\kappa}\in(-1,0)$ and hence $\tilde{\kappa}$ can be used to
parametrize the solution for $a$ and $b$. After some algebra, we find that the
exact solution of equations (\ref{8}) and (\ref{13}) parametrized in terms of
$\tilde{\kappa}$ is given by
\begin{align}
a^{3}(\tilde{\kappa})  & =\frac{4M\kappa_{0}}{3}\,\left\vert \tilde{\kappa
}\right\vert \left(  1-\tilde{\kappa}^{4}\right)  \,\left(  \frac
{1+\tilde{\kappa}^{2}}{1+3\tilde{\kappa}^{2}}\right)  ^{2},\label{15}\\
b^{3}(\tilde{\kappa})  & =\frac{9M}{2\kappa_{0}^{2}\tilde{\kappa}^{2}%
}\,\left(  1-\tilde{\kappa}^{2}\right)  \left(  1+3\tilde{\kappa}^{2}\right)
.\label{16}%
\end{align}
Using (\ref{14}) to express $\kappa$ in terms of $x$, we can easily verify
that in the far exterior limit $x\rightarrow\infty$, $\tilde{\kappa}%
^{2}\rightarrow0$ and the above solution tends to (\ref{5}), describing a
black hole of mass $M$. On the other hand, deep inside the black hole at
$x\rightarrow-\infty$, $\tilde{\kappa}^{2}\rightarrow1$ and we obtain
asymptotic (\ref{6}) corresponding to the de Sitter space with $H=\kappa
_{0}/3$. Thus, the obtained exact solution smoothly matches the desired
asymptotics, in agreement with our general consideration above. The function
$a$ reaches its maximum at $\tilde{\kappa}_{\ast}=-1/\sqrt{5}$. The horizons,
which occur at $a=1,$ exist only if $a\left(  \tilde{\kappa}_{\ast}\right)
\geq1.$ This condition is satisfied only if
\begin{equation}
M\geq M_{\min}=\frac{5^{5/2}}{18\kappa_{0}}.\label{17}%
\end{equation}
Otherwise, no horizon exists and the solution (\ref{14}), (\ref{15}),
(\ref{16}) describes solitonic-like objects whose metric is completely static
and approaches the de Sitter metric in the center. For black holes with the
minimum mass $M_{\min}$, there is only one horizon with vanishing surface
gravity and hence these minimal black holes represent the stable remnants of
evaporating black holes.

One can check that solution (\ref{14}), (\ref{15}), (\ref{16}) satisfies
$a^{2}=\dot{b}^{2}$ and hence the hypersurfaces $T=\textup{const.}$ are
exactly spatially flat, in complete agreement with the assumption under which
it was derived. To better understand the properties of this solution, it is
more illuminating to go back to the familiar singular static coordinates
(\ref{1}). From $\left(  \ref{4}\right)  $ it follows that $dr=adx$ and
therefore $a^{2}=\dot{b}^{2}$ implies that $b=r$ everywhere. Setting $b=r$ in
equation ( \ref{16}), we obtain an algebraic equation for $\tilde{\kappa
}\left(  r\right)  $, which can be solved perturbatively and the obtained
result can be substituted in (\ref{15}) to determine $a^{2}\left(  r\right)  $
in the \textquotedblleft static coordinates\textquotedblright,\ where the
metric is given by (\ref{1}).

For $r\rightarrow\infty$ where $\tilde{\kappa}^{2}\ll1$ we find the expansion
\begin{equation}
1-a^{2}=1-\frac{2M}{r}\left[  1-\mathcal{O} \left( \left(  \frac{r_{\ast}}%
{r}\right)  ^{3}\right) \right] ,\label{19}%
\end{equation}
where $r_{\ast}=\left(  144M/5\kappa_{0}^{2}\right)  ^{1/3}$ is the
\textquotedblleft radial\textquotedblright\ coordinate at which $\tilde
{\kappa}_{\ast}=-1/\sqrt{5}$ and the curvature becomes comparable to the
limiting curvature. For large black holes with $M\gg M_{\min}$ the outer
horizon, defined by $a\left(  r_{+}\right)  =1,$ is located at
\begin{equation}
r_{+}=2M\left[  1-\mathcal{O}\left(  \left(  \frac{M_{\min}}{M}\right)
^{2}\right) \right] ,\label{20}%
\end{equation}
and the deviations from the Schwarzschild solution become significant only
deeply \textquotedblleft inside\textquotedblright\ the black hole at $r\ll
r_{+}.$ Close to the limiting curvature $\left(  \tilde{\kappa}^{2}%
\rightarrow1\right)  ,$ that is for $r\ll r_{\ast}, $ the metric is well
approximated by
\begin{equation}
1-a^{2}=1-(Hr)^{2}\left[  1-\mathcal{O}\left( \left(  \frac{r}{r_{\ast}%
}\right)  ^{3}\right) \right] \label{21}%
\end{equation}
where $H=\kappa_{0}/3$ and the inner de Sitter horizon occurs at
\begin{equation}
r_{-}=H^{-1}\left[  1+\mathcal{O}\left( \frac{M_{\min}}{M}\right)  \right]
.\label{22}%
\end{equation}
If the mass $M$ is comparable to the minimal mass $M_{\min}$, these two
asymptotics fail to describe the region close to the horizons. In the minimal
case $M=M_{\min}$ inner and outer horizon coincide. Expanding $a\left(
\tilde{\kappa}\right)  $ around its maximum at $\tilde{\kappa}_{\ast}$ and
using $\left(  \ref{16}\right)  $ to express $\tilde{\kappa}$ in terms of $r$,
we find that the near horizon metric of such a minimal black hole is given by
\begin{equation}
1-a^{2}\approx\frac{10}{7}\left(  1-\frac{r}{r_{\ast}}\right)  ^{2},\label{23}%
\end{equation}
where $r_{+}=r_{-}=r_{\ast}=2\sqrt{5}/\kappa_{0}$. Note the similarity to the
near horizon metric of an extremal charged Reissner-Nordstr{\"o}m black hole.

\emph{Black hole thermodynamics.---}
The Hawking temperature $T_{H}$ is determined by the surface gravity
(\ref{11}) at the exterior horizon $x_{+}.$ For solution $\left(  15\right)  $
we find%
\begin{equation}
T_{H}=\frac{g_{s}}{2\pi}=\frac{\kappa_{0}}{6\pi}\,\left\vert \tilde{\kappa
}_{+}\right\vert \,\frac{1-5\tilde{\kappa}_{+}^{2}}{1+3\tilde{\kappa}_{+}^{2}%
},\label{24}%
\end{equation}
where $\tilde{\kappa}_{+}=\tilde{\kappa}\left(  x_{+}\right)  \in(-1/\sqrt
{5},0)$. Since $a\left(  \tilde{\kappa}_{+}\right)  =1$, we can use (\ref{15})
to express $M$ also through $\tilde{\kappa}_{+}$ as
\begin{equation}
M=\frac{3}{4\kappa_{0}\left\vert \tilde{\kappa}_{+}\right\vert \left(
1-\tilde{\kappa}_{+}^{4}\right)  }\left(  \frac{1+3\tilde{\kappa}_{+}^{2}%
}{1+\tilde{\kappa}_{+}^{2}}\right)  ^{2}.\label{25}%
\end{equation}
The formulae (\ref{24}) and (\ref{25}) implicitly define the relation
$T_{H}\left(  M\right) $. In particular, at large mass we reproduce in leading
order the familiar Hawking formula
\begin{equation}
T_{H}=\frac{1}{8\pi M}\left[  1+\mathcal{O}\left(  \left(  \frac{M_{\min}}%
{M}\right)  ^{2}\right)  \right]  .\label{25a}%
\end{equation}
Instead of diverging as $M\to0$, the temperature reaches its maximum value
$T_{\max}\sim10^{-2}\kappa_{0}$ at $\left\vert \tilde{\kappa}_{+}\right\vert
\approx0.23$ which corresponds to $M=M_{c}\approx1.32M_{\min}$. At this point
the negative heat capacity diverges and becomes positive for $M<M_{c}$. Close
to the minimal mass the temperature decreases as
\begin{equation}
T_{H} \propto\sqrt{M-M_{\min}}.
\end{equation}
According to the Stefan-Boltzmann law, the rate of energy loss of a radiating
body is determined by $\frac{\mathrm{d}M}{\mathrm{d}t}\propto-T_{H}^{4} A$
where $A=4\pi r_{+}^{2}$ is the horizon area. For an evaporating black hole
close to minimal mass $A\sim M_{\min}^{2}$, and hence it will eventually
approach $M_{\min}$ according to $M(t)-M_{\min}\propto t^{-1}$. That is, the
final product of black hole evaporation is a stable minimal remnant with
$M=M_{\min}$ and vanishing Hawking temperature.

Finally, taking into account that the Bekenstein entropy of the black hole is
equal to $S=A/4 = \pi r_{+}^{2}=\pi b^{2}\left(  \tilde{\kappa}_{+}\right)  $
it is straightforward to verify that the modified first law of thermodynamics
becomes
\begin{equation}
G(\tilde{\kappa}_{+})\mathrm{d}M=T_{H}\,\mathrm{d}S,\label{26}%
\end{equation}
where $G(\tilde{\kappa}_{+})=f^{-1}(\tilde{\kappa}_{+})$ is the value of the
gravitational constant at the outer horizon.

\emph{Conclusions.---}
We have shown that a modification of classical Einstein theory at very high
curvatures, implementing asymptotic freedom at limiting curvature, can spare
us from having to deal with non-perturbative quantum gravity (at least in
application to cosmological and black hole problems). The existence of a
sub-Planckian limiting curvature at which the gravitational constant vanishes
can resolve the black hole singularity and replace it with a patch of de
Sitter space (similar to \cite{MMFrolov}). Moreover, in this theory the final
result of black hole evaporation are remnants whose near horizon geometry is
similar to the extremal Reissner-Nordstr{\"{o}}m geometry. Therefore, the
Hawking temperature of these remnants is equal to zero. In distinction from
extremal Reissner-Nordstr{\"{o}}m black holes, they do not exhibit a
singularity and, because they have no charge, they are stable.  From the
maximal extension of the solution obtained above (cf. \cite{Note1}), it
becomes clear that these remnants can store an unlimited amount of
information. This information lies in the absolute future of external
observers and remains forever inaccessible for them. Hence, their degeneracy
should not lead to any paradoxes in calculating physical processes observed by
external observers. This suggests one of the possible ways for a resolution of
the information paradox (see \cite{UnruhWald}). Moreover, the stable remnants
can serve as well as Dark Matter candidates.

\begin{acknowledgments}
The work of A. H. C is supported in part by the National Science Foundation
Grant No. Phys-1518371. The work of V.M. and T.B.R. is supported by the
Deutsche Forschungsgemeinschaft (DFG, German Research Foundation) under
Germany's Excellence Strategy -- EXC-2111 -- 390814868.
\end{acknowledgments}

\bibliographystyle{plain}
\bibliography{bib}

\end{document}